\documentclass[aps,prl,twocolumn,groupedaddress,showpacs,preprintnumbers,amsmath,amssymb]{revtex4}
\usepackage{psfig}
\begin{document}

\title{Towards a quantitative phase-field model of two-phase solidification}

\author{R. Folch}
\altaffiliation[Present address: ]
{Universiteit Leiden, Postbus 9506, 2300 RA Leiden, The Netherlands.}
\author{M. Plapp}

\affiliation{Laboratoire de Physique de la Mati\`ere Condens\'ee,
CNRS/\'Ecole Polytechnique, 91128 Palaiseau, France}


\begin{abstract}
We construct a diffuse-interface model of two-phase solidification 
that quantitatively reproduces the classic free boundary
problem on solid-liquid interfaces in the thin-interface limit.
Convergence tests and comparisons with boundary integral simulations 
of eutectic growth
show good accuracy for steady-state lamellae,
but the results for limit cycles depend
on the interface thickness through the trijunction behavior.
This raises the fundamental
issue of diffuse multiple-junction dynamics.
\end{abstract}

\pacs{64.70.Dv, 81.30.Fb, 05.70.Ln}

\maketitle

Complex microstructures that arise during alloy solidification are a 
classical example of pattern formation \cite{patterns}
and influence the mechanical properties of the finished material \cite{Kurz}.
A long-standing challenge is to understand the pattern selection starting
from the basic ingredients: bulk transport, solute and heat
rejection on the solidification front, and the front's local response. 
Simple as it may seem, this free boundary problem (FBP) accurately
describes many experimental features,
but has few analytic solutions, so that
numerical modeling is mandatory.

The phase-field method \cite{pf} has become the method of choice for 
simulating solidification fronts \cite{solidif}, and more generally for 
tackling FBPs and interfacial pattern formation phenomena, e.g.
in materials science \cite{materials} and fluid flow \cite{vf}.
Its main advantage 
(essential in three dimensions) 
is that it 
circumvents front tracking by using {\it phase fields} to locate 
the fronts. These fields interpolate between different constant values 
in each bulk phase through interfacial regions of thickness $W$.
The model is then required to reproduce the FBP
in the sharp-interface limit, 
in which the extra length scale $W$ vanishes.

In practice, simulations have to resolve the variation 
of the phase fields through the interfaces, so that 
$W$ must stay finite. Their
results generally depend on the ratio $W/\ell$, 
where $\ell$ is a relevant length scale of the FBP. 
Explicit corrections to the original FBP to first order
in $W/{\ell}$ have been calculated by a so-called
{\em thin-interface} analysis in a few cases,
and some canceled out \cite{KarmaRappel,Almgren,KarmaPRL,vf}.
A complete cancellation,
achieved for single-phase solidification \cite{KarmaRappel,KarmaPRL},
means that results become independent of $W/\ell$
for some finite value of $W$. The correct FBP is then reproduced
already at that value, much larger than the thickness of real interfaces,
enabling {\em quantitative} contact in three dimensions
between simulations, theory, and experiments 
in {\em reasonable} simulation times \cite{KarmaTip}.

Here, we extend these advances to two-phase
solidification, which already includes
the most widespread solidification microstructures after
dendrites: eutectic composites. They consist of
alternate lamellae of two solids ($\alpha$ and $\beta$) 
or of rods of one solid embedded in the other, 
growing from a melt $L$ near a eutectic 
point, where all three phases coexist
at equilibrium.
The interplay between capillarity and diffusive bulk transport
between adjacent solid phases can give rise to more complex patterns
and nonlinear phenomena such as bifurcations, limit cycles,
solitary waves, and spatiotemporal chaos \cite{Ginibre97}.

A two-phase solidification front consists of (i) solid-liquid
interfaces and (ii) trijunction points where all three phases meet.
Our strategy is to construct a phase-field model that 
allows us to analyze the thin-interface behavior 
of (i) separately from (ii).
We quantitatively reproduce the correct FBP on (i); (ii)
satisfy Young's law at equilibrium.
We test convergence in $W/\ell$ for lamellar eutectic growth at
experimentally relevant parameters, and compare our results 
to boundary integral (BI) \cite{KarmaSarkissian} simulations
and other phase-field models.
For steady states, we achieve good agreement with the BI and a 
drastically improved, fast convergence compared to previous models.
In contrast, convergence is slow for limit cycles,
due to a trijunction behavior affecting the overall dynamics.

We use one phase field $p_i$ to indicate presence ($p_i=1$)
or absence ($p_i=0$) of each phase $i=\alpha,\beta,\rm L$ 
in the spirit of volume fractions \cite{Steinbach96}, which requires
\begin{equation}
p_\alpha + p_\beta + p_{\rm L} = 1.
\label{constraint}
\end{equation}
The phase fields evolve in time to minimize a free energy functional 
${\cal F}$ of $\vec p\equiv(p_\alpha,p_\beta,p_{\rm L})$, the solute
concentration, and temperature,
\begin{equation}
\frac{\partial p_i}{\partial t} =-\frac{1}{\tau(\vec p)}
\left.\frac{\delta{\cal F}}{\delta p_i}\right|_{p_\alpha + p_\beta + p_{\rm L} = 1} \;\forall i,
\label{modelforp}
\end{equation}
where $\tau(\vec p)$ is a phase-dependent relaxation time.
This classical problem of minimizing a functional
subject to a constraint is treated by the method of Lagrange multipliers;
$(\delta{\cal F}/\delta p_i)|_{p_\alpha + p_\beta + p_{\rm L} = 1} =
\delta{\cal F}/\delta p_i - (1/3)\sum_j \delta{\cal F}/\delta p_j$
for three phases,
where the functional derivatives on the r.h.s. 
are now taken as if all $p_i$ were independent.

To distinguish between phases, earlier phase-field models of 
two-phase solidification 
used either the usual solid--liquid phase field and 
the local concentration \cite{Karma94} or introduced
a second,  $\alpha$--$\beta$ phase field \cite{Wheeler96}.
Across a solid--liquid interface, both fields must vary, 
so that their dynamics are coupled, which complicates a thin-interface analysis.
The same is true for a generic choice of $\cal F$ in 
Eq.~(\ref{modelforp}). However, if 
on an $i$--$j$ interface we can assure that the third
phase field $p_k$ is exactly zero, $p_i$ or $p_j$ can be 
eliminated using Eq.~(\ref{constraint}), so that
the interface can be described in terms of a single
independent variable. This was recently achieved using
a free energy with cusp-like minima \cite{Garcke99}, 
but no thin-interface analysis is available for that model.
We also achieve absence of the third phase, but using a {\em smooth} 
free energy, by requiring $p_k=0$ to be a stable solution 
for $p_k$ of Eqs. (\ref{modelforp}) for each $i$--$j$ interface:
\begin{subequations}
\label{valleys}
\begin{eqnarray}
\label{flatness}
\left. \frac{\delta {\cal F}}{\delta p_k} 
\right |_{p_\alpha + p_\beta + p_{\rm L} = 1, \,p_k=0} = 0 \;\;\forall k, \\
\label{convexity}
\left. \frac{\delta^2 {\cal F}}{\delta p_k^2} 
\right |_{p_\alpha + p_\beta + p_{\rm L} = 1, \,p_k=0} > 0 \;\;\forall k.
\end{eqnarray}
\end{subequations}
The advantage is that the simplest choice for ${\cal F}$ yields a model
that turns out to coincide with the quantitative model of Ref.~\cite{KarmaPRL}
on those $i$--$j$ interfaces.  

To construct our free energy, we split it into parts, 
\begin{equation}
\label{functional}
{\cal F} = \int_V f_{\rm grad} + f_{\rm TW} + \tilde{\lambda} f_c.
\end{equation}
The first is a free energy penalty
\begin{equation}
f_{\rm grad} = \frac{W^2}{2} \sum_i  \left |\vec\nabla p_i\right |^2
\end{equation}
for the gradients of the phase fields that provides the 
interface thickness $W$.
The next is a triple-well potential
\begin{equation}
f_{\rm TW} = \sum_i p_i^2\left(1-p_i\right)^2
\end{equation}
that generates the basic ``landscape'':
one well per pure phase and ``valleys'' with double-well 
profiles along each $p_k=0$ cut, separated
by a potential barrier on trijunctions
$p_\alpha=p_\beta=p_{\rm L}=1/3$. 
The last part
has a strength $\tilde{\lambda}$ (a constant that controls convergence) and
couples the phase fields $p_i$
to the temperature $T$ and the solute concentration $C$ through 
$c(C) \equiv (C-C_{\rm E})/\Delta C$, where
$\Delta C\equiv C_\beta-C_\alpha$, $C_\alpha$ and $C_\beta$
are the limits of the eutectic plateau, and 
$(C_{\rm E},T_{\rm E})$ is the eutectic point,
\begin{equation}
\label{fc}
f_c=\sum_i g_i(\vec p) \left [B_i(T)-\mu A_i(T)\right],
\end{equation}
where we have introduced the chemical-potential-like variable
$\mu \equiv c - \sum_i A_i(T) h_i$,
and $g_i(\vec p)$ and $h_i(\vec p)$ (given below)
interpolate between $0$ for $p_i=0$ and $1$ for $p_i=1$.

The term $f_c$ drives the system out of equilibrium 
by unbalancing the pure phase free energies:
Each well $i$ is shifted by an amount $B_i-\mu A_i$.
The equilibrium value $\mu = \mu_{\rm eq}^{ij} = (B_j-B_i)/(A_j-A_i)$
gives equal shifts and hence restores the balance between phases 
$i$ and $j$; from the definition of $\mu$, we obtain
$c_i^{ij} = A_i+\mu_{\rm eq}^{ij}$ for
the concentration in phase $i$ coexisting with phase $j$.
A eutectic phase diagram with constant concentration 
gaps and straight liquidus and solidus lines is generated by
$A_i=c_i\equiv c(C_i)$ and $B_i = c_i(T-T_E)/(m_i\Delta C)$, 
with $m_i$ the (signed) liquidus slopes, $i=\alpha,\beta$.
Non-constant concentration gaps and peritectic 
phase diagrams can also be treated. 
Without loss of generality, $A_{\rm L}=B_{\rm L}=0$.

In order for $\mu = \mu_{\rm eq}^{ij}$ to keep the balance all across the 
$i$--$j$ interface as $p_i$ goes from $0$ to $1$, we require
\begin{equation}
\label{antisymmetry}
g_i(p_i,p_j,0)=1-g_i(p_j,p_i,0) \; \forall i.
\end{equation}
Otherwise, several thin-interface corrections arise \cite{Almgren,KarmaPRL}.
The simplest choice satisfying also Eq.~(\ref{flatness}) is
$g_i = p_i^2 \{15(1-p_i) [1+p_i-(p_k-p_j)^2] + p_i (9p_i^2-5) \} /4$.

The evolution of $\mu$ is obtained from its definition and mass 
conservation, $\partial_t c + \vec\nabla\cdot\vec J = 0$,
$\vec J = - Dp_{\rm L}\vec\nabla \mu + \vec J_{\rm AT}$:
\begin{equation}
\label{modelformu}
\frac{\partial \mu}{\partial t} = 
D \vec\nabla\cdot \left (p_{\rm L}\vec\nabla\mu\right )
-\sum_i A_i \frac{\partial h_i}{\partial t}
-\vec\nabla\cdot\vec J_{\rm AT},
\end{equation}
where $-Dp_{\rm L}\vec\nabla\mu$ is the usual diffusion current,
with a diffusivity that varies from $D$ in the liquid
to $0$ in the solid (one-sided model), and  $\vec J_{\rm AT}$
is an extension of the antitrapping current introduced
in~\cite{KarmaPRL} that counterbalances spurious solute 
trapping,
\begin{equation}
\label{antitrapping}
\vec J_{AT} \equiv - \hat n_{\rm L} \frac{W}{2\sqrt{2}}
\sum_{i=\alpha,\beta} 
A_i \frac{\partial p_i}{\partial t}(\hat n_i \cdot\hat n_{\rm L}),
\end{equation}
where $\hat n_i = -\vec\nabla p_i/|\vec\nabla p_i|$ 
are unit vectors normal to $i$--L interfaces, and 
$\hat n_i \cdot\hat n_{\rm L}$ prevents solute exchange
between the two solids.
The model is not variational, because of the
term $\vec J_{\rm AT}$ and because $\mu \neq \partial f_c / \partial c$,
but enables us to use $h_i=p_i$, which
allows for a coarser discretization~\cite{KarmaRappel}.

Our model [Eqs. (\ref{modelforp}) and (\ref{modelformu})]
has stable interface solutions connecting two coexisting phases
$i$ and $j$: 
$\mu = \mu_{\rm eq}^{ij}$,
$p_i=1-p_j=\{1\pm \tanh[r/(W\sqrt{2})]\}/2$
(with $r$ the distance to the interface), $p_k = 0$.
Since these solutions are identical for all $i$-$j$ pairs, 
so are the $i$-$j$ surface tensions.
Unequal surface tensions can be obtained by adding new
terms in Eq.~(\ref{functional}) that shift the $i$--$j$
free energy barriers.

Remarkably, on solid--liquid ($i$--L) interfaces, 
assuming a weak dependence of the $A_i$, $B_i$ on $T$, 
and $\tau(\vec p) = \tau_i$, the change of variables
$\phi_i = p_i - p_{\rm L}$, $u=(\mu^{i \rm L}_{eq}-\mu)/A_i$
maps Eqs. (\ref{modelforp}) and (\ref{modelformu})
to the quantitative model with constant concentration gap 
in~\cite{KarmaPRL}, up to numerical prefactors.
The thin-interface limit can hence be deduced by inspection and
yields the classic FBP on $i$--L interfaces,
\begin{subequations}
\label{fbp}
\begin{eqnarray}
\label{diffusion}
\partial_t c & = & D \nabla^2 c, \\
\label{stefan}
-D \hat n_i \cdot \vec\nabla c & = & v_n (c_i^{i L}-c_L^{i L}), \\
\label{gt}
c & = & \mp \left(\frac{T-T_{\rm E}}{|m_i|\Delta C} + d_i \kappa + \beta_i v_n \right ),
\end{eqnarray}
\end{subequations}
where Eq.~(\ref{diffusion}) holds in the liquid and the others are boundary 
conditions on the interface that has normal velocity $v_n$ and curvature
$\kappa$; the minus (plus) refers to  $i= \alpha$  ($\beta$),
and the capillary lengths $d_i$ and kinetic coefficients $\beta_i$
read in terms of our model parameters
\begin{eqnarray}
\label{capillary}
d_i & = & a_1 \frac{W}{|A_i|\tilde{\lambda}}, \\
\label{kinetic}
\beta_i & = & a_1 
   \left [ \frac{\tau_i}{|A_i|\tilde{\lambda} W}
	- a_2\frac{|A_i|W}{D} \right ],
\end{eqnarray}
with $a_1=\sqrt{2}/3$ and $a_2=1.175$. The constant $\tilde\lambda\propto W/d_i$ in
Eqs.~(\ref{functional}), (\ref{capillary}) and (\ref{kinetic})
controls the convergence to the original 
FBP. Any set of $\beta_i$ can be treated with suitable $\tau_i$. 
We consider here $\beta_\alpha=\beta_\beta=0$,
which is achieved with
$\tau_i=a_2 A_i^2 \tilde{\lambda}W^2 / D$.
The different $\tau_i$ for $A_\alpha\neq A_\beta$ 
(e.g. different concentration gaps) are interpolated by
$\tau(\vec p) = \bar\tau + (1/2)(\tau_\alpha - \tau_\beta)
(p_\alpha - p_\beta)/(p_\alpha+p_\beta)$,
$\tau(p_\alpha+p_\beta=0)=\bar\tau$, 
with $\bar\tau=(\tau_\alpha+\tau_\beta)/2$.

We test our model in directional solidification with
$T = T_{\rm E} + G (z-Vt)$, where $G>0$ is the thermal 
gradient and $V>0$, the pulling speed, both directed along
the $z$ axis. Half a eutectic lamellae pair of total width $\lambda$ is simulated 
in two dimensions ($x$ and $z$) with no-flux boundary conditions in the midline
of each lamella, using a finite-difference Euler scheme with a grid spacing
$\Delta x=0.8W$ (coarser far into the liquid to improve efficiency).
We adopt $l_D/{\bar d}=51200$ and ${\bar l}_T/l_D=4$,
where $l_D \equiv D/V$ is the diffusion length,
$l_T^i \equiv |m_i| \Delta c/G$ are the thermal lengths,
and $\bar d \equiv (d_\alpha + d_\beta)/2$, 
$\bar l_T \equiv (l_T^\alpha+l_T^\beta)/2$.
These correspond to typical experimental values
$G\approx 100 K/cm$, $V\approx 1\mu m/s$ for 
CBr$_{\rm 4}$-C$_{\rm 2}$Cl$_{\rm 6}$, an organic eutectic for
which accurate experimental data exist \cite{Ginibre97}.
We use $m_\alpha=-m_\beta$, $c_\alpha=-c_\beta$ (a symmetric
phase diagram) or $m_\beta/m_\alpha=-2$,
$-c_\beta/c_\alpha=d_\alpha/d_\beta=2.5$ (one close to 
CBr$_{\rm 4}$-C$_{\rm 2}$Cl$_{\rm 6}$). In both
cases $\mu(z\rightarrow+\infty)=0$ (eutectic composition). 
We test convergence to the thin-interface limit with decreasing $W$
by conversely increasing $\lambda/W$ 
while keeping all the ratios above and $\lambda/\lambda_{\rm min}$ fixed,
where $\lambda_{\rm min}\propto \sqrt{\bar{d}l_D}$ 
is the minimal undercooling spacing \cite{Jackson66}. 
This is achieved by varying the constant 
$\tilde\lambda$ in Eq.~(\ref{capillary}).

\begin{figure}
\centerline{\psfig{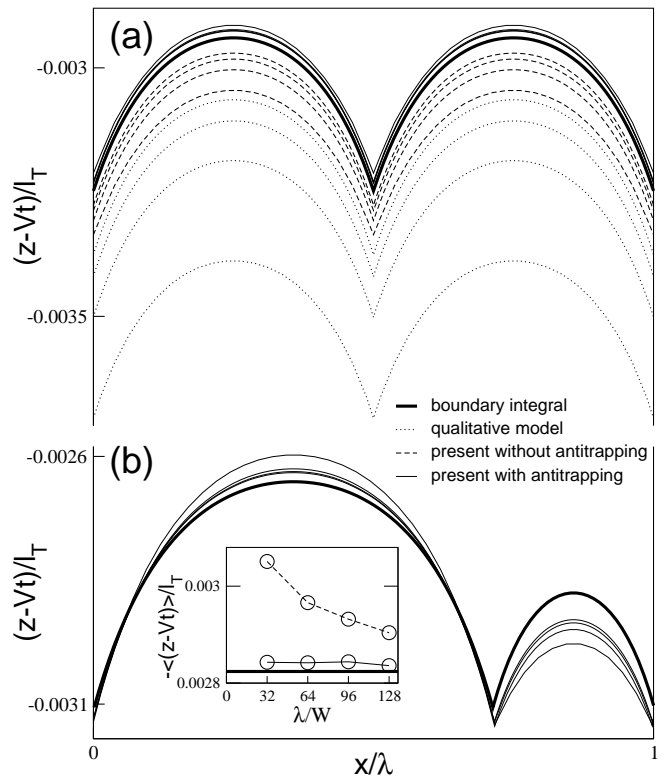}}
\caption{\label{figsteady}
Steady-state lamellae pair profiles 
(dimensionless undercooling {\it vs.} $x/\lambda$)
for different models. 
Four curves at $\lambda/W=$32, 64, 96 and 128 shown per model;
curves closer to the boundary integral: larger  $\lambda/W$.
[$\lambda/W=$ 64--128 collapse for the present model with
antitrapping current in (a)].
Phase diagram  used: (a) symmetric; (b) close to CBr$_4$--C$_2$Cl$_6$.
See parameters in the text.
Inset: Averaged undercooling in (b) {\it vs.}  $\lambda/W$,
compared to that without antitrapping current.
}
\end{figure}

Figure \ref{figsteady} shows the solid--liquid interfaces of a 
steady-state lamellae pair calculated by different phase-field
models and the boundary integral method 
(BI) \cite{KarmaSarkissian}
for $\lambda\approx \lambda_{\rm min}$.
For the symmetric phase diagram [Fig.  \ref{figsteady}(a)],
our model (thin solid lines) agrees well with the BI (thick solid line). 
Moreover, the curves at
$\lambda/W=64$, 92 and 128 are indistinguishable. 
This means that the results are independent
of $\lambda/W$ for $\lambda/W\geq 64$, the signature of a 
quantitative model. In contrast, if we remove
the antitrapping current in our model, $\vec J_{\rm AT}=\vec 0$, 
which leads to solute trapping and finite interface kinetics,
the results depend on $\lambda/W$ for all the range
from 32 (bottom dashed line) to 128 (top one). The
convergence of models not backed by a 
thin-interface analysis can even be slower,
as shown by the dotted curves
for a qualitative version of our model with $h_i=g_i$
violating Eq. (\ref{antisymmetry}) and 
$\vec J_{\rm AT}=\vec 0$ \cite{Dresden}; in this situation,
several thin-interface corrections to the FBP occur 
simultaneously \cite{Almgren,KarmaPRL}.

Results are similar for the phase diagram close to 
CBr$_{\rm 4}$-C$_{\rm 2}$Cl$_{\rm 6}$ [Fig. \ref{figsteady}(b)]. 
The convergence is somewhat slower, since one
of the lamellae is thinner and needs to be properly resolved.
Some small deviation from the BI persists, probably due to the 
trijunction behavior (see below). In the inset, we plot the 
average undercooling vs. $\lambda/W$. 
This is a less stringent test, as shown by the fact that 
results for our model are converged already for $\lambda/W=32$. 
However, those for the model  with $\vec J_{\rm AT}=\vec 0$ still depend
on $\lambda/W$ at $\lambda/W=128$, which illustrates
how all corrections need to be canceled
before quantitative results can be achieved.

\begin{figure}
\psfig{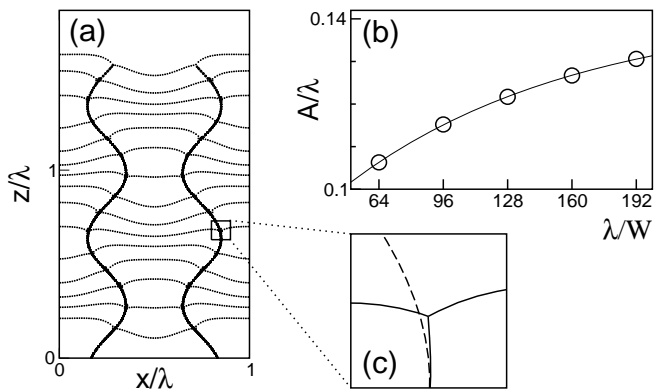}
\caption{\label{figcycle} Limit cycles. 
(a) Superimposed snapshots of the interfaces 
at constant time intervals for $\lambda/W=64$. 
Thicker lines: $\alpha$--$\beta$ interfaces.
(b) Amplitude of the trijunction oscillation 
in units of $\lambda$ {\it vs}. $\lambda/W$. 
The line is a fit that yields $A(\lambda/W\rightarrow\infty)/\lambda=0.142$.
(c) Blowup of $6.4W\times 6.4W$. Solid lines: trijunction passage;
dashed line: later $\alpha$--$\beta$ interface.}
\end{figure}

Next, we increase $\lambda$
to $\approx 2.2 \lambda_{\rm min}$, close above the threshold
$\lambda\approx 2\lambda_{\rm min}$ \cite{KarmaSarkissian} for 
the bifurcation from steady lamellae to oscillatory limit cycles, 
a situation in which the oscillation amplitude is very sensitive
to all parameters. Indeed, for the symmetric phase diagram and 
$\lambda/W=64$, the qualitative model of Ref.~\cite{Dresden} still 
yields lamellae, whereas the present model correctly produces 
cycles, which are shown in Fig. \ref{figcycle}(a). However,
the amplitude of the trijunction oscillation $A/\lambda$, 
defined as its maximal displacement in $x/\lambda$, 
strongly depends on $\lambda/W$, as shown in Fig. \ref{figcycle}(b).
An extrapolation yields $A(\lambda/W\rightarrow\infty)=0.142\lambda$,
not far from the BI result $A=0.139\lambda$, 
but the results are still not converged for $\lambda/W=192$,
in strong contrast to the steady-state behavior.
This suggests that some correction(s) to the FBP in
$W/\lambda$ remain in our model. 
Since solid--liquid interfaces are controlled, 
we turn to the the trijunctions.

The solid (dashed) lines in Fig.~\ref{figcycle}(c) show a first 
(later) snapshot of the interfaces close to a turning point
of the trijunction trajectory. In the later one the trijunction 
has moved away and only the $\alpha$--$\beta$ interface remains,
which has slightly moved sideways. 
In the one-sided FBP, (i) the $\alpha$--$\beta$ interface {\em cannot} move,
so it is the trace left by the trijunction, and (ii) its 
direction close to the trijunction
approaches that of the trijunction velocity.
In a diffuse-interface model, the diffusivity {\em behind} the
trijunction point $p_\alpha=p_\beta=p_{\rm L}=1/3$ 
falls to zero on the scale of $W$, so that 
(i) and (ii) do not hold. We consistently observe the displacement 
to be a fraction of $W$ fairly independent of $\lambda/W$, and the 
whole trijunction to be slightly rotated with respect to its velocity, 
features also observed for the steady state in Fig.~\ref{figsteady}b.
This effect
explains the remaining mismatch between phase-field and BI in
Fig.~\ref{figsteady}b and the slow convergence of $A/\lambda$ here.

We have presented a phase-field model of two-phase
solidification that coincides with the best models to date
\cite{KarmaRappel,KarmaPRL} on 
solid--liquid interfaces,
whose dynamics are completely controlled.
This has allowed us to identify the role of
diffuse trijunctions in the convergence of the results. 
Understanding their dynamics is both a fundamental issue and
a prerequisite for a fully quantitative modeling of 
multiphase solidification:
First, a thin-interface analysis of the 
trijunction region in the phase-field model is lacking.
Even so, our model
is expected to be precise and yield a substantial
efficiency gain for small curvatures of trijunction 
trajectories, which makes it a promising tool for
three-dimensional simulations. 
Second,
the free boundary problem to converge
to should also be reconsidered.
It was shown elsewhere that Young's condition 
on the angles {\em between} interfaces 
is violated out of equilibrium for
kinetically limited growth \cite{Caroli};
here, the {\em global} trijunction rotation 
was found to be fairly independent
of the interface thickness, so that it might
persist for real nanometric interfaces.
These effects 
should be further investigated,
possibly by atomistic simulations.

We thank S. Akamatsu and G. Faivre for discussions,
A. Karma for the BI code, and 
Centre National d'\'Etudes Spatiales (France) 
for support.
R. F. also acknowledges a European Community Marie Curie Fellowship.

\end{document}